\documentclass[sigconf]{acmart}
\usepackage{graphicx,verbatim,multirow,dcolumn}
\usepackage{amsmath,amssymb,bm}
\usepackage{amsthm}
\usepackage{booktabs}
\usepackage{comment}
\usepackage{lettrine}
\usepackage{xcolor}
\usepackage{array}
\usepackage{graphicx}
\usepackage{microtype}
\usepackage{multirow}
\usepackage{float}
\usepackage{tikz}
\usepackage{bm}
\usepackage{url}
\usepackage[normalem]{ulem}
\usepackage[aboveskip=1pt,skip=1pt,belowskip=1pt]{caption}
\usepackage{subcaption}
\usepackage[capitalise]{cleveref}

\copyrightyear{2019} 
\acmYear{2019} 
\setcopyright{acmlicensed}
\acmConference[WSDM '19]{The Twelfth ACM International Conference on Web Search and Data Mining}{February 11--15, 2019}{Melbourne, VIC, Australia}
\acmBooktitle{The Twelfth ACM International Conference on Web Search and Data Mining (WSDM '19), February 11--15, 2019, Melbourne, VIC, Australia}
\acmPrice{15.00}
\acmDOI{10.1145/3289600.3290976}
\acmISBN{978-1-4503-5940-5/19/02}

\newcommand{\phantomsubfigure}[1]{\begin{subfigure}[b]{0.1\textwidth}\phantomcaption\label{#1}\end{subfigure}}
\newcommand{\xhdr}[1]{\vspace{0.75mm}\noindent{{\bf #1.}}\hspace{0.5mm}}

\newcommand{\ffrac}[2]{#1 \big/ #2}

\newtheorem{theorem}{Theorem}

\newcolumntype{C}[1]{>{\centering\let\newline\\\arraybackslash\hspace{0pt}}m{#1}}

\begin{document}

\title{Random Spatial Network Models with Core-Periphery Structure}


\author{Junteng Jia}
\orcid{1234-5678-9012}
\affiliation{%
  \institution{Cornell University}
  \streetaddress{107 Hoy Road}
}
\email{jj585@cornell.edu}

\author{Austin R. Benson}
\orcid{1234-5678-9012}
\affiliation{%
  \institution{Cornell University}
  \streetaddress{107 Hoy Road}
}
\email{arb@cs.cornell.edu}

\date{\today}

\begin{abstract}
  Core-periphery structure is a common property of complex networks, which is a
  composition of tightly connected groups of core vertices and sparsely connected
  periphery vertices. This structure frequently emerges in traffic systems,
  biology, and social networks via underlying spatial positioning of the
  vertices. While core-periphery structure is ubiquitous, there have been limited
  attempts at modeling network data with this structure. Here, we develop a generative,
  random network model with core-periphery structure that jointly accounts for
  topological and spatial information by ``core scores'' of vertices. Our model
  achieves substantially higher likelihood than existing generative models of
  core-periphery structure, and we demonstrate how the core scores can be used
  in downstream data mining tasks, such as predicting airline traffic and
  classifying fungal networks. We also develop nearly linear time algorithms for
  learning model parameters and network sampling by using a method akin to the fast multipole
  method, a technique traditional to computational physics, which allow us to
  scale to networks with millions of vertices with minor tradeoffs in
  accuracy.
\end{abstract}

\maketitle

\section{Network core-periphery structure}\label{sec:intro}
Networks are widely used to model the interacting components of complex systems
emerging from biology, ecosystems, economics, and
sociology~\cite{Newman_2010,Easley_2010,Barabasi_2016}.
A typical network consists of a set of vertices V and a set of edges E,
where the vertices represent discrete objects (e.g., people or cities) and
the edges represent pairwise connections (e.g., friendships or
highways).
Networks are often described in terms of local properties such as vertex degree
or local clustering coefficients and global properties such as diameter or the
number of connected components.
At the same time, a number of mesoscale proprieties are consistently
observed in real-world networks, which often reveal important structural
information of the underlying complex systems; arguably the most well-known is
community structure, and a tremendous amount of effort has been devoted to its
explanation and algorithmic
identification~\cite{Schaeffer_2007,Andrea_2009,Fortunato_2010,Moore_2017}.

\begin{figure}[t]
  \centering
  \phantomsubfigure{fig:com_cp_A}
  \phantomsubfigure{fig:com_cp_B}
  \phantomsubfigure{fig:com_cp_C}
  \phantomsubfigure{fig:com_cp_D}
\includegraphics[width=0.92\linewidth]{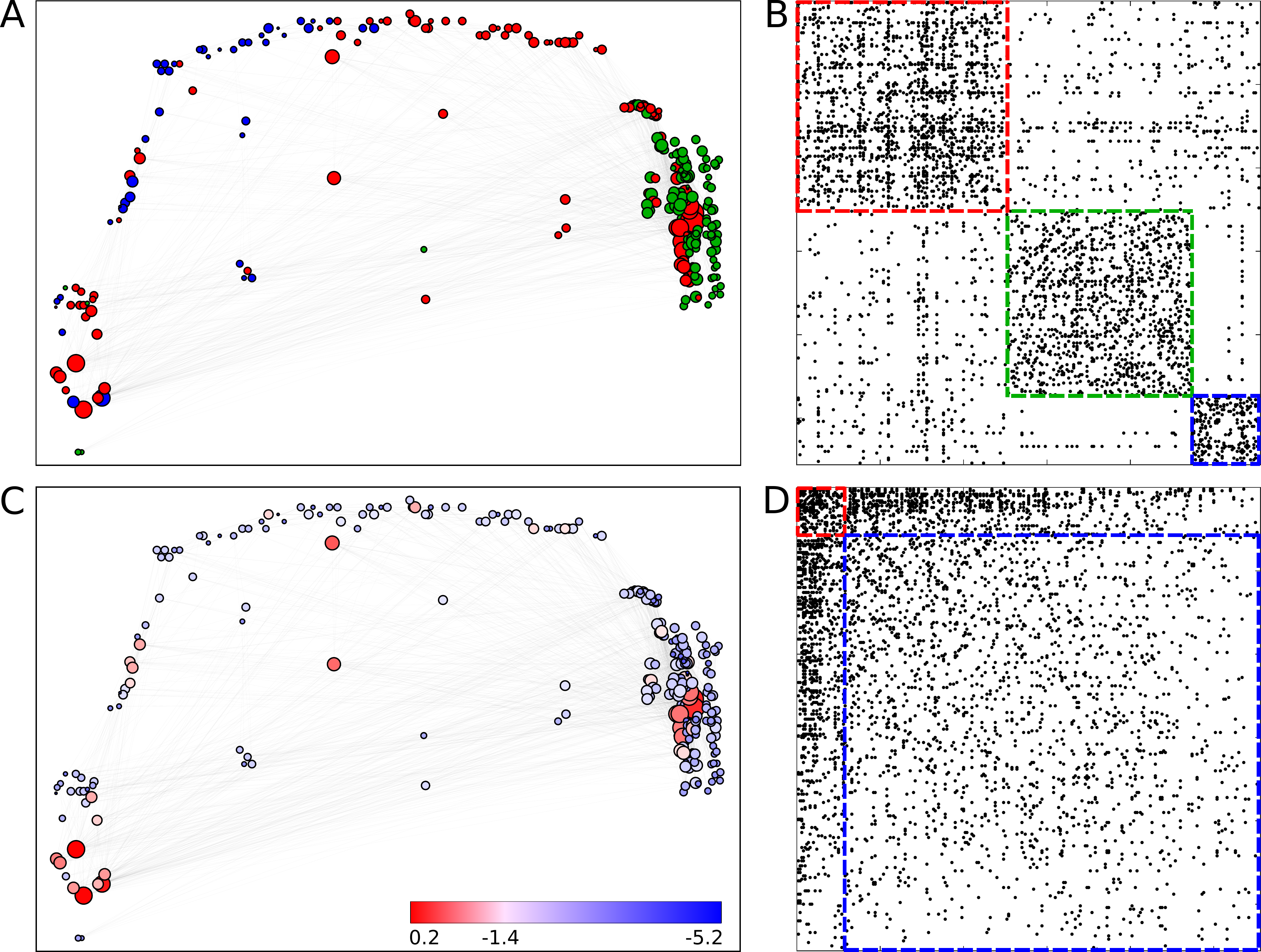}
\caption{%
Community (A--B) and core-periphery (C--D) structure in the \textit{C. elegans}
network, where vertices are neurons, and edges are neural
connections. Vertex coordinates are neuron locations in the lateral plane~\cite{Varier_2011} and
vertex sizes are proportional to the square root of degrees.
(A) The Louvain algorithm~\cite{Blondel-2008-louvain} finds three
communities (identified here by the three colors).
(B) The adjacency matrix ordered by the three communities.
(C) Our proposed model learns vertex ``core scores'' based on spatial location
and connectivity, where larger core scores are more indicative of a vertex being
in the core; here, the maximum and minimum core scores are 0.2 and -5.2.
(D) The adjacency matrix ordered by decreasing vertex core score.
Vertices with high core scores are both spatially
distributed and densely connected.
}
\label{fig:com_cp}
\end{figure}

Another important mesoscale structure is \emph{core-periphery} structure, 
although such structure has received relatively little attention.
In contrast to community detection, which separates vertices into 
several well-connected modules, core-periphery identification involves
finding sets of cohesive core vertices and periphery vertices
loosely connected to both each other and the cores.
This type of structure is common in
traffic~\cite{Rombach_2017,Rossa_2013},
economic~\cite{Rossa_2013,Boyd_2010},
social~\cite{Holme_2005,Zhang_2015,Cucuringu_2016},
and biological~\cite{Silva_2008}
networks.
Oftentimes, spatial information is a driving factor in the core-periphery structure~\cite{Wellhofer_1989,Varier_2011,Verma_2016}.
For example, in the \emph{C. elegans} neural network shown in \cref{fig:com_cp},
most long-distance neural connections are between tightly connected early-born
neurons, which serve as hubs and constitute the core~\cite{Varier_2011}.
Identifying core-periphery structure not only provides us with a new perspective to
study the mesoscale structure in networked systems but can also leads to
insights on their functionalities~\cite{Verma_2014}.

At the same time, random network models are useful for analyzing and understanding
networks~\cite{Newman_2002_rg,leskovec2007graph}.
For example, they are routinely used as the null model to verify that
features of real world networks are not due 
randomness~\cite{Middendorf_2005,Goldenberg_2010,Yin_2018} and also serve
to identify community structure, as is the case with
the stochastic block model and its variants~\cite{Holland_1983,airoldi2008mixed,peixoto2017bayesian}
as well as methods such as BigCLAM~\cite{yang2013overlapping} and CESNA~\cite{Yang2013}.
These models successfully incorporate community structure, but have yet
to effectively model core-periphery structure.
While block modeling can also be used to identify core-periphery
structure~\cite{Zhang_2015}, we later show that such an approach is limited.
Moreover, in general, spatial information is not often incorporated into random network
models, even though it can play a large role in their structure.

Here, we present a random network model that generates networks with
core-periphery structure.
In our model, each vertex has a real-valued \textit{core score} to reflect its
role in the core.
Our model assumes the edges are generated by a random process and the
probability that two vertices connect is an increasing function of their core scores.
Given an arbitrary network, we infer the model parameters (core scores)
by maximum likelihood estimation.
We prove that at any local optimum of the likelihood function, the expected
degree of any node under the model is the same as the given network.
Therefore, our model can be an alternative to the Chung-Lu model for generating
random networks~\cite{Aiello_2001}.

Our model also accounts for spatial locations of vertices (if such data is
available).
Spatial networks emerge appear in application such as trade,
transportation, the power grid, and the Internet, where there is a cost
associated with the length of the edges~\cite{Barthelemy_2011} and are known to
carry core-periphery structure~\cite{Rossa_2013,Lee_2014,Dong_2015}.
In such cases, topology alone does not explain many proprieties in spatial networks,
such as ``small world'' navigability.
Our model accounts for spatial information by decreasing the probability of an edge between a pair of vertices
with their distance.
We show that at local optima of our likelihood function, the log of the expected
geometric mean edge length of networks generated by our model is the same as in
the given network from which the parameters are learned.

Spatial information enables us to design efficient algorithms for maximizing
likelihood and generating random networks.
The main idea is that if a set $S$ of vertices are far away in space from some
given vertex $u$, then the effect of $S$ on $u$ can be efficiently approximated.
We perform this approximation in a hierarchical manner similar to the fast multipole
method, a ``top 10 algorithm of the twentieth century''~\cite{cipra2000best}.
Although the algorithm is traditionally used to accelerate $N$-body simulations
in physics, we adapt it here to develop nearly linear time algorithms for
likelihood optimization and network sampling. This lets us scale to
networks with millions of vertices using a commodity server.

Our model has substantially higher likelihood compared to the two other random
graph models that explicitly incorporate core-periphery structure.
%
%
We also show that the learned core scores are useful for downstream data mining
and machine learning tasks; specifically, learned core scores out-perform
existing baselines for predicting airport traffic and classifying fungal
networks.


\section{Model and basic inference}\label{sec:method}
In this section, we develop our model for generating networks with
core-periphery structure and a straightforward maximum likelihood procedure to
learn model parameters for a given input network.
We analyze the basic properties of the model without worrying about computation
(efficient algorithms are the focus of \cref{sec:algo}).
We provide two technical results about local optima of the likelihood function
for our model: (i) the expected degree of each vertex matches the input
network and (ii) the expected aggregated log-distance matches the
input network.

\subsection{A generative core-periphery model} \label{subsec:model}
\xhdr{Basic model}
In our basic model, we start with $n$ vertices,
where each vertex $u$ has a real-valued core score $\theta_{u}$.
For every pair of vertices $u$ and $v$, we add an edge between them with probability
\begin{align}
\textstyle \rho_{uv} = \ffrac{e^{\theta_{u} + \theta_{v}}}{(e^{\theta_{u} + \theta_{v}} + 1)}.
\label{eq:rho_uv_basic}
\end{align}
As a sanity check, the edge probability $\rho_{uv} \in [0, 1]$ and increases
monotonically as a function of the combined core score.
Thus, vertices that are both ``in the core'' (i.e., have large core scores) are more likely to connect.
Two special cases have significant meaning.
First, if all vertices have the same core score $\theta_{0}$, then the
model is the Erd\H{o}s-R\'{e}nyi model with edge probability
$p = e^{2\theta_{0}} / (e^{2 \theta_{0}} + 1)$.
Second, if the vertices are partitioned into a core $V_{c}$ with core score
$\theta_{0}$ and a periphery $V_{p}$ with core score $-\theta_{0}$,
then as $\theta_{0}$ increases, the model converges to a traditional
block model for social networks: the core vertices $V_{c}$
form a clique while the periphery vertices $V_{p}$ form an independent set,
and every pair of vertices $u \in V_{c}$ and $v \in V_{p}$ is connected
with probability 0.5.

\xhdr{Full model}
Now we incorporate spatial information into the model.
All of our subsequent algorithms and analysis are then presented for this
model, which includes the basic model in \cref{eq:rho_uv_basic} as a special case.
Our model incorporates spatial information by adding a kernel function $K_{uv}$
to the denominator of edge probability
\begin{align}
\textstyle \rho_{uv} = \ffrac{e^{\theta_{u} + \theta_{v}}}{(e^{\theta_{u} + \theta_{v}} + K_{uv}^{\epsilon})}.
\label{eq:rho_uv_general}
\end{align}

The kernel function $K_{uv}$ can be any arbitrary non-negative function provided
by the user, but we mainly focus on metric kernels to study spatial
networks where core-periphery structure is prevalent.
We will introduce kernel functions as we go through examples, but a
simple kernel is Euclidean distance:
$K_{uv} = \|\bm{x}_{u} - \bm{x}_{v}\|_{2}$,
where $\bm{x}_{u}$ and $\bm{x}_{v}$ are the spatial positions of vertices $u$ and $v$.
We also include a tuning parameter $\epsilon$ to control the ``nearsightedness'' of
vertices.
When $\epsilon=0$, the edge probability $\rho_{uv}$ is independent of the edge
length, and we recover the basic model.
As $\epsilon$ increases, the fraction of long distance edges in the generated
networks steeply decrease.
While the parameter $\epsilon$ could be baked into the kernel, we find
it useful to optimize $\epsilon$ in conjunction with the core scores
$\bm{\theta}$ when fitting the model to network data (see \cref{subsec:inference}).

\xhdr{Relationship to small worlds}
Our model is inspired in part by the Kleinberg navigable small-world model,
where random edges are added to a lattice
network with probability inversely proportional to powers of
distance~\cite{Kleinberg_2000_navigation,Kleinberg_2000_algorithm}.
In real-world networks that are sparse, we expect the edge probability
$\rho_{uv}$ in the corresponding generalized model to often be small.
In those cases, the edge probability in our model can be
approximated by
\begin{align}
\textstyle \rho_{uv} \approx \ffrac{\rho_{uv}}{(1-\rho_{uv})} = \ffrac{e^{\theta_{u} + \theta_{v}}}{K_{uv}^{\epsilon}}.
\label{eq:rho_uv_general_approx}
\end{align}
Our  model thus has an interpretation for social networks:
actors live somewhere in space and have different social statuses; people
have a higher chance to connect if they are closer, and individuals with
higher social status are likely to have more acquaintances.

\subsection{Inference via Likelihood Maximization} \label{subsec:inference}
For inference, we take as input an undirected network and a kernel function and
output a set of real-valued vertex core scores and a real-valued
tuning parameter $\epsilon$ for the kernel function.
To do so, we maximize the log-likelihood $\Omega$:
\begin{align}
\textstyle \Omega = \sum_{u<v} \left[A_{uv} \log \rho_{uv} + (1-A_{uv}) \log (1-\rho_{uv})\right].
\label{eq:omega}
\end{align}
The gradient of the edge probability with respect to the model parameters is given by
simple closed-form expressions:
%
%
\begin{align}
\textstyle \frac{\partial \rho_{uv}}{\partial \theta_{u}} = \rho_{uv} (1 - \rho_{uv}), \hspace{0.10in} \frac{\partial \rho_{uv}}{\partial \epsilon} = -\rho_{uv}(1-\rho_{uv}) \cdot \log K_{uv}.
\label{eq:drho_dpara}
\end{align}

First, we focus on the derivatives of the objective function with respect to the core scores:
\begin{align}
\textstyle \frac{\partial \Omega}{\partial \theta_{w}}
&= \textstyle \sum_{u<v} \frac{\partial \rho_{uv}}{\partial \theta_{w}} \left(\frac{A_{uv}}{\rho_{uv}} - \frac{1-A_{uv}}{1-\rho_{uv}}\right) 
= \textstyle \sum_{u<v} (\delta_{wu} + \delta_{wv})(A_{uv} - \rho_{uv}) \nonumber \\
&= \textstyle \sum_{u \neq w} A_{wu} - \sum_{u \neq w} \rho_{wu}.
\label{eq:domega_dtheta}
\end{align}
Here, $\delta_{ab}$ is the Kronecker delta function.
Note that $\sum_{u \neq w} A_{wu}$ is the degree of vertex $w$ in the given
network, while $\sum_{u \neq w} \rho_{w u}$ is the expected degree of
vertex $w$ in the model.
This observation implies our first theorem about the learned model.
\begin{theorem}\label{thm:degree}
  At any local maximizer of the objective function $\Omega$ with respect to the
  core scores $\bm{\theta}$, the expected degree of every vertex in the random
  model equals the degree in the input network.
\end{theorem}

Next, we look at the derivative of the objective function with respect to the tuning parameter $\epsilon$,
\begin{align}
\textstyle \frac{\partial \Omega}{\partial \epsilon} &= \textstyle \sum_{u<v} \frac{\partial \rho_{uv}}{\partial \epsilon} \left(\frac{A_{uv}}{\rho_{uv}} - \frac{1-A_{uv}}{1-\rho_{uv}} \right) \nonumber \\
&=\textstyle -\sum_{u<v} \log K_{uv} \cdot \rho_{uv}(1-\rho_{uv}) \cdot \left(\frac{A_{uv}}{\rho_{uv}} - \frac{1-A_{uv}}{1-\rho_{uv}} \right) \nonumber \\
&= \textstyle -\sum_{u<v} A_{uv} \log K_{uv} + \sum_{u<v} \rho_{uv} \log K_{uv}.
\label{eq:domega_depsilon}
\end{align}
Let $\sum_{u<v} A_{uv} \log K_{uv}$ be the \textit{aggregated log-distance}, which measures the overall edge length.
Then $\sum_{u<v} \rho_{uv} \log K_{uv}$ is the expected aggregated log-distance in networks generated by the learned model.
This observation implies our next result.
\begin{theorem}
  At any local maximizer of the objective function $\Omega$ with respect to
  $\epsilon$, the expected aggregated log-distance in the random model equals
  the true aggregated log-distance in the input network.
\label{thm:aggregated_log_distance}
\end{theorem}
\noindent In other words, optimizing the tuning parameter $\epsilon$ forces the
overall distances in the model to match the original network.

Next, define the log geometric mean edge length (log-GMEL) of a network as
\begin{align}
\textstyle \textrm{log-GMEL} 
= \log \left[\prod_{u<v \in E} K_{uv}\right]^{\frac{1}{\lvert E \rvert}} 
= \frac{1}{\lvert E \rvert}\sum_{u<v \in E}\log K_{uv}
\label{eq:geom_mean_edge_length}
\end{align}
We argue that the log-GMEL of a random network is close to the log-GMEL of the
input graph as well.
By the law of large numbers, in the limit of
large networks, the number of edges $\lvert E \rvert$
model concentrates around its expectation.
When the number of edges is sharply concentrated about its expectation, we can
approximate the expected log-GMEL by
$\mathbf{E}\left[\textrm{log-GMEL}\right] \approx \ffrac{\mathbf{E}\left[\sum_{u<v \in E} \log K_{uv}\right]}{\mathbf{E}\left[\lvert E \rvert\right]}$,
which is the expected aggregated log-distance divided by the expected number of
edges.
According to \cref{thm:degree,thm:aggregated_log_distance}, the expected number
of edges and the expected aggregated log-distance equal to those in the input
network, respectively.
Thus, the expected log-GMEL should roughly be the log-GMEL of the input network.
In \cref{sec:data}, we validate this numerically.

Since the derivatives of the objective function can be evaluated analytically,
we use a gradient-based approach to optimize the likelihood.
However, the log-likelihood objective $\Omega$ is not necessarily convex.
Therefore the computed optimal set of parameters $\{\bm{\theta}^{*}, \epsilon^{*}\}$
is not guaranteed to be the global maximizer of $\Omega$.
Finally, even though spatial networks are our primary focus in this paper, 
\cref{thm:degree} still holds for the basic model in \cref{eq:rho_uv_basic}.
Thus, the basic model can be used for both core-periphery structure
detection and as an alternative to the Chung-Lu model for generating random
networks with arbitrary sequences of expected degrees.

With a naive implementation of model inference, evaluating the objective
function (\cref{eq:omega}) or gradient
(\cref{eq:domega_dtheta,eq:domega_depsilon}) takes $\mathcal{O}(\lvert V
\rvert^{2})$ time, regardless of the choice for the kernel function.
Similarly, if we sample a random network by determining the connectivity of
every pair of vertices sequentially, the overall cost is also
$\mathcal{O}(\lvert V \rvert^{2})$, even if the generated network is sparse.
In the next section, we design nearly linear-time approximation algorithms for
both model inference and network generation when the kernel function is a metric
that satisfies the triangle inequality.


\section{Fast Algorithms}\label{sec:algo}
The quadratic scaling of the naive algorithm limits the model's
applicability to large-scale networks.
In order to resolve this problem, we use a method akin to the fast multipole
method (FMM) to exploit structural sparsity in the computations when the kernel
is a metric, which results in nearly linear time algorithms by sacrificing a
controlled amount of accuracy.
The main idea of the approach is that the joint influence of a group of vertices
$S$ on a far away group of vertices $T$ can be well
approximated for metric kernels.

\subsection{Efficient Model Inference}
We will use a gradient-based method to optimize the objective function, but the
computational bottleneck of optimizing model parameters is the evaluation of the
objective function and its gradient.
Here, we take advantage of the fact that the expressions for the objective
function and its gradient are in close analogy to the gravitational potential
and forces in $N$-body simulations.
Similar to the gravitational potential, the objective function $\Omega$ as well
as its derivative $\partial \Omega / \partial \epsilon$ consists of
$\mathcal{O}(\lvert V \rvert^{2})$ pairwise interactions which decay as a function of
distance (assuming a metric kernel).
Furthermore, our objective function and derivative also accumulate contributions
with respect to the core scores over all pairs of vertices.
These similarities motivate us to use ideas from the FMM to exploit
the ``structural sparsity'' in our computations.

\xhdr{Background on the FMM}
The FMM is a numerical algorithm for accelerating computation of $N$-body
simulations, which require the (approximate) accumulation of $O(N^2)$
pairwise interactions~\cite{rokhlin1985rapid,greengard1987fast}.
In physics, these calculations look like $P(u)=\sum_{v \in S}k(u,v)f(v)$, where
$P$ is the potential (e.g., gravitational potential), $k$ is the kernel (e.g.,
$k(u, v) = 1 / \| \bm{x}_{u} - \bm{x}_{v} \|_{2}$ in gravitational potential),
and $f$ is a weight function, and we want $P(u)$ for all $u \in S$.
The key conceptual idea is that the interaction between two groups of
well-separated particles can be well-approximated by a \emph{single} interaction
between the total mass of the two groups. This gives rise to the ``structural
sparsity'' of the problem.
The main mathematical idea is to use a (multipole) expansion of the kernel function to
approximate these well-separated interactions~\cite{ying2012pedestrian,beatson1997short}.

\begin{figure}[t]
\centering
\includegraphics[width=0.87\linewidth]{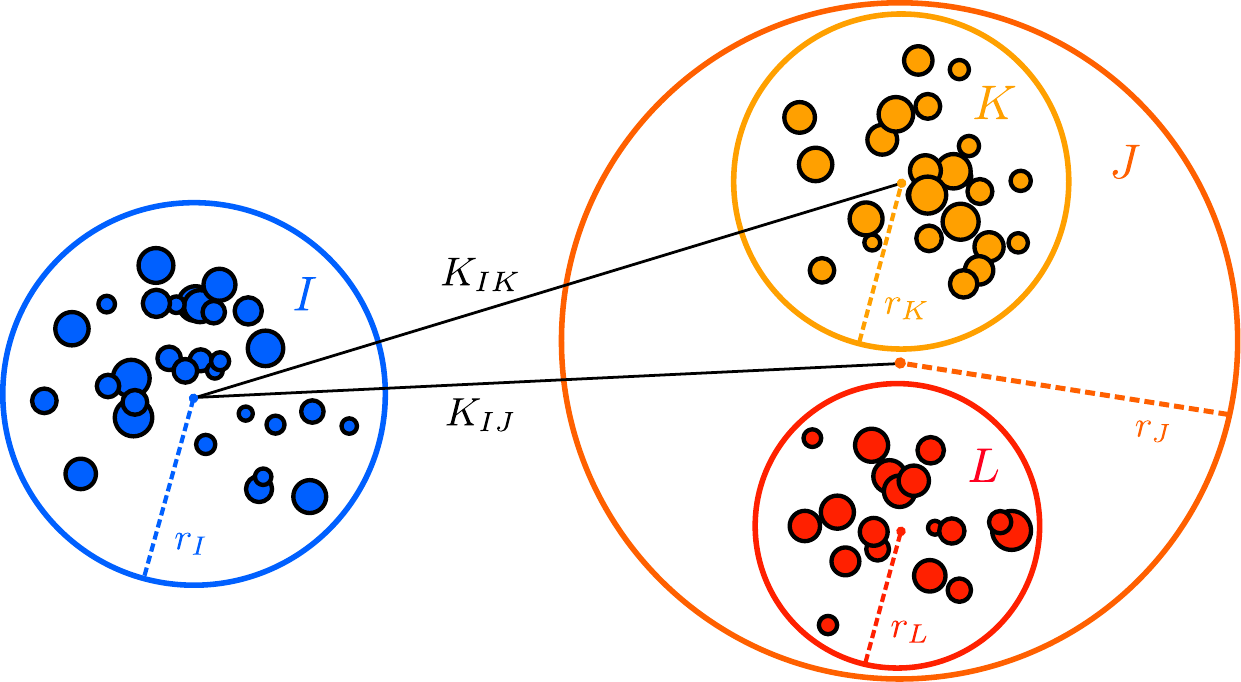}
\caption{%
  Interactions between vertices in $I$ and $J$ with our algorithm.
  If the accuracy criteria in \cref{eq:fmm_criterion_1,eq:fmm_criterion_2} are
  satisfied, then we approximate the interaction between all vertices in $I$ and
  all vertices in $J$ with a single point in each ball (\cref{eq:omega_fmm_approx}).
  Otherwise, we recurse and consider the interactions between $I$ and $K$ as
  well as $I$ and $L$.
}
\label{fig:fmm_split}
\end{figure}

\begin{figure}[t]
\centering
\includegraphics[width=0.69\linewidth]{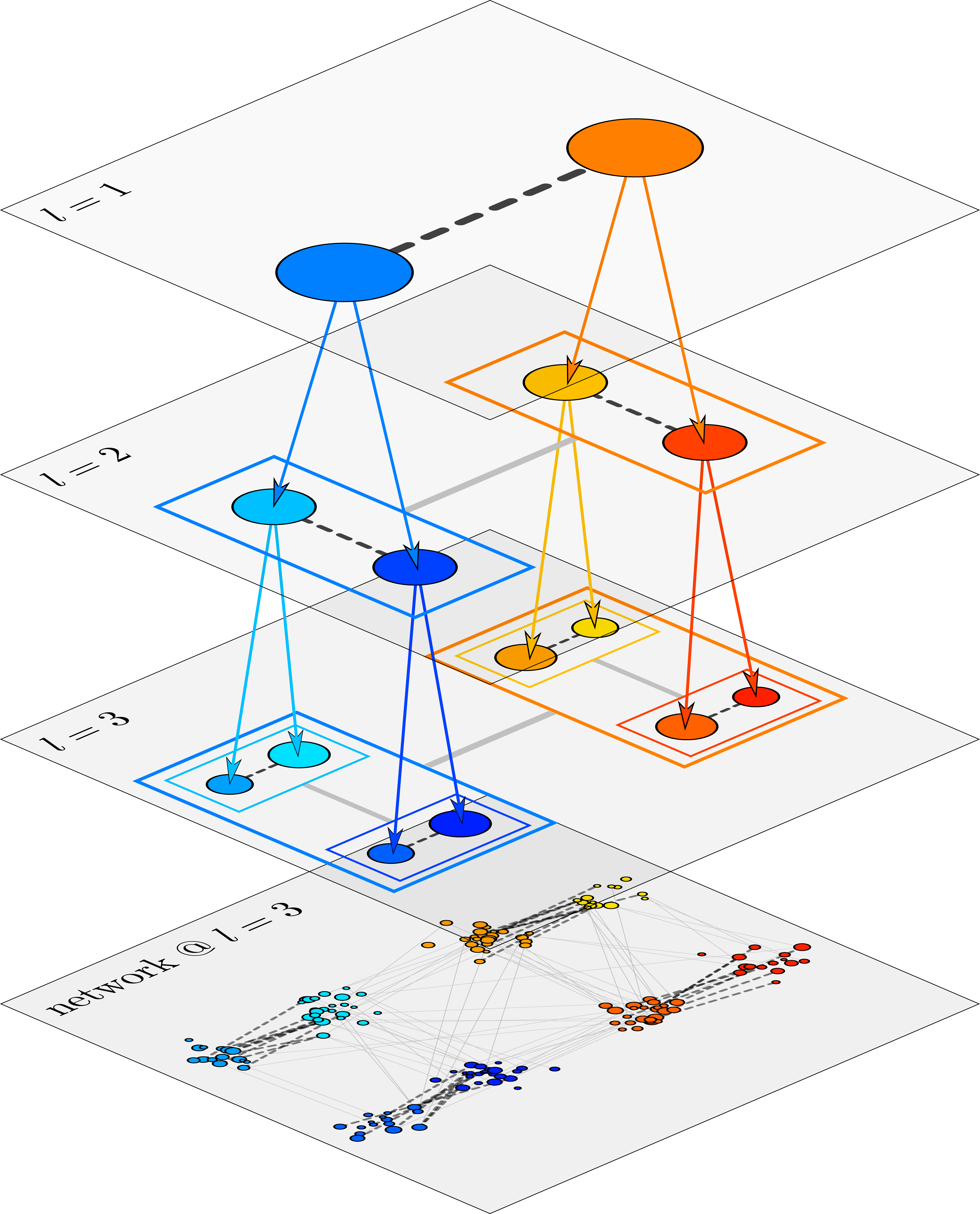}
\caption{
The top three levels in the metric-tree data structure. Each circle in the
diagram represents a metric-ball that encapsulates spatially adjacent vertices.
This data structure is used for efficient evaluation of the objective function
and its gradient as well as for efficient network generation.
For objective function evaluation, the dashed lines at each layer of the diagram
represent the pairwise interactions accumulated at the corresponding level of
the metric-tree.
For random network generation, the dashed lines represent the edges
sampled at each level of the metric-tree between sibling metric-balls, while the
solid lines represent edges that are already generated in previous levels.
The bottom layer of the diagram shows the random network generated at the $l=3$
level of the metric tree; edges between vertices within the same metric ball at this
level are not yet considered.
}
\label{fig:algo}
\end{figure}

We use use a hierarchical metric-tree decomposition of the spatially distributed vertices.
The root of the tree (level $l=0$) is a metric-ball containing all vertices, and
we recursively bisect the network to fit into smaller metric-balls at lower
levels (see \cref{fig:algo,fig:fmm_split}).
For example, each node at the $l=1$ level of metric-tree represents a
metric-ball that encaptulates half of the vertices in the network, which is
further divided into two child nodes at the $l=2$ level.
Leaf nodes in the metric-tree represent a metric-ball that contains only
one vertex from the network. We are using the term {\em node} to refer to
metric-balls in this data structure---not to actual vertices in a network; we
use {\em vertex} when referring to graphs.

\xhdr{Evaluating the objective function}
In order to exploit the ``separation'' of far-away interactions,
we re-write the objective in \cref{eq:omega} to
first separate the vertex pairs that are connected:
\begin{align*}
\textstyle \Omega &= \textstyle \sum_{u<v} \left[A_{uv} \log \rho_{uv} + (1-A_{uv}) \log (1-\rho_{uv})\right] \nonumber \\
       &= \textstyle \sum_{u<v \in E} \log \rho_{uv} + \sum_{u<v \not\in E} \log(1 - \rho_{uv}) \nonumber \\
       &= \textstyle \sum_{u<v \in E} \log \frac{\rho_{uv}}{(1 - \rho_{uv})} - 
          \sum_{u<v} \log(1 + \ffrac{e^{\theta_{u}+\theta_{v}}}{K_{uv}^{\epsilon}}).
\end{align*}

Recall from \cref{thm:degree} that at any local maximizer of the objective
function, the expected degree of a vertex in the model equals its
degree in the given network (i.e., $\sum_{u \neq w}\rho_{wu}=\sum_{u \neq w}A_{wu}$).
If the given network is sparse, then $\sum_{u \neq w} A_{wu} \ll \lvert V \rvert$ holds
for most vertices, $\rho_{uv} \ll 1$ holds for most vertex pairs,
and $z_{uv} \equiv \ffrac{e^{\theta_{u} + \theta_{v}}}{K_{uv}^{\epsilon}} = \ffrac{\rho_{uv}}{(1-\rho_{uv})} \ll 1$.
In other words, the learned core scores in the given network are small enough
that $e^{\theta_{u} + \theta_{v}} \ll K_{uv}$ for most pairs of vertices.
Thus, we can use the Maclaurin expansion for $\log(1+z_{uv})$ to approximate the
objective function,
\[
  \textstyle \Omega \approx 
  \sum_{u<v \in E} \log \frac{\rho_{uv}}{1 - \rho_{uv}} - 
                \sum_{u<v} \sum_{t=1}^{T} \frac{(-1)^{t-1}}{t}\left(\frac{e^{\theta_{u}+\theta_{v}}}{K_{uv}^{\epsilon}}\right)^{t}.
\]
Here, $T$ is a small constant ($T = 4$ in our implementation; larger expansions
provided little benefit in accuracy).
The first term in the expansion only sums over connected vertex pairs and can be calculated in $\mathcal{O}(\lvert E \rvert)$ time.
The summation over the $\mathcal{O}(\lvert V \rvert^{2})$ pairwise interactions
in the second term of the expansion is accelerated by grouping vetices in the
same metric-ball and computing interactions between two groups of vertices at
once.
While we have made several approximations in our arguments, our numerical
experiments in \cref{subsec:experiment_fmm} show that they are valid on
real-world networks.

Our FMM-like implementation is similar to the algorithm presented in the
original literature known as the ``tree-code''~\cite{Apple_1985}.
At each level $l$ of the metric-tree, for every pair of sibling metric-balls
$I,J$, we sum over the interaction between every pair of vertices with one end
in $I$ and the other in $J$,
\begin{align}
\textstyle \sum_{u<v} \left(\frac{e^{\theta_{u}+\theta_{v}}}{K_{uv}^{\epsilon}}\right)^{t} 
= \sum_{l} \sum_{\substack{I<J \in B_{l}\\ p(I)=p(J)}} \left[\sum_{u \in I, v \in J} \left(\frac{e^{\theta_{u}+\theta_{v}}}{K_{uv}^{\epsilon}}\right)^{t}\right],
\label{eq:omega_fmm}
\end{align}
where $B_{l}$ represents the set of metric-balls at level $l$ of the
metric-tree, and $p(I),p(J)$ denote the parent metric-balls of $I,J$
respectively.

In order to guarantee the accuracy of the FMM algorithm, the pairwise
interactions between vertices in $I$ and $J$ (in the square bracket of
\cref{eq:omega_fmm}) can only be computed at once if the separation between
the metric-balls is large relative to their radii, i.e.,
\begin{align}
\textstyle \ffrac{K_{IJ}}{(r_{I}+r_{J})} > \delta_{1},
\label{eq:fmm_criterion_1}
\end{align}
where $r_{I},r_{J}$ are the radii of the metric-balls, $K_{IJ}$ measures the
metric distance between their centers of mass, and $\delta_{1}$ is a user-input
accuracy.
For our problem, we also have to make sure the assumption $z_{uv} \ll 1$
holds for every pair of vertices with one end in $I$ and the other in $J$, i.e.,
\begin{align}
\textstyle e^{ \max\{\theta_{u} \;\vert\; u \in I\} } \cdot e^{ \max\{\theta_{v} \;\vert\; v \in J\} } / K_{IJ} < \delta_{2},
\label{eq:fmm_criterion_2}
\end{align}
where $\delta_{2}$ is another user-input accuracy parameter.
We use $\delta_{1}=2.0$ and $\delta_{2}=0.2$ as default in our implementation. However, the
user can tune these parameters to trade off accuracy and computation time (we
conduct trade-off experiments in \cref{subsec:experiment_fmm}).

If the two accuracy criteria in \cref{eq:fmm_criterion_1} and
\cref{eq:fmm_criterion_2} are satisfied, then the long-range pairwise
interactions between $I$ and $J$ can be approximated as follows:
\begin{align}
  \textstyle \sum_{u \in I, v \in J} \left(\frac{e^{\theta_{u}+\theta_{v}}}{K_{uv}^{\epsilon}}\right)^{t} \approx
  \ffrac{\left(\sum_{u \in I} e^{\theta_{u}}\right)^t\left(\sum_{v \in J} e^{\theta_{v}}\right)^t}{K_{IJ}^{t \epsilon}}.
\label{eq:omega_fmm_approx}
\end{align}
On the other hand, if one or more accuracy criteria are not satisfied, then
without loss of generality, $I$ is the metric-ball with smaller radius
($r_{I} < r_{J}$), and we compute the interactions between metric-ball $I$ and each
of the two child metric-balls of $J$ separately (\cref{fig:fmm_split}).
The exact computational complexity of the FMM algorithm depends on the parameters
$\delta_{1}$ and $\delta_{2}$. 
In practice, the objective function is approximated to within $1\%$ error using the default
parameters (see \cref{subsec:experiment_fmm}), and the overall time complexity is $\mathcal{O}(\lvert E \rvert+\lvert V \rvert\log\lvert V \rvert)$.

\xhdr{Evaluating the gradient}
We can also use the FMM to approximate the derivative of $\Omega$
with respect to the core score $\theta_{w}$.
At each level $l$ of the metric tree, we sum over the interactions between
vertex $w$ and all the vertices in the metric-ball $J_{l}$ which is the sibling
to the metric-ball $I_{l}$ that contains $w$,
\begin{align}
\textstyle \frac{\partial \Omega}{\partial \theta_{w}} 
&= \textstyle \sum_{u \neq w} A_{wu} - \sum_{u \neq w} \ffrac{e^{\theta_{u} + \theta_{w}}}{\left[e^{\theta_{u} + \theta_{w}} + K_{uw}^{\epsilon}\right]} \nonumber \\
&= \textstyle \sum_{u \neq w} A_{wu} - \sum_{l} \sum_{u \in J_{l}} \ffrac{e^{\theta_{u} + \theta_{w}}}{\left[e^{\theta_{u} + \theta_{w}} + K_{uw}^{\epsilon}\right]} \nonumber \\
& \textstyle \approx \sum_{u \neq w} A_{wu} - \ffrac{e^{\theta_{w}}\sum_{l, u \in J_{l}} e^{\theta_{u}}}{\left[\frac{e^{\theta_{w}}}{\lvert J_{l}\rvert}\sum_{u \in J_{l}} e^{\theta_{u}} + K_{I_{l}J_{l}}^{\epsilon}\right]}.
\label{eq:domega_dtheta_fmm}
\end{align}

The first term of \cref{eq:domega_dtheta_fmm} is the degree of vertex $w$, which
can be evaluated upfront in $\mathcal{O}(\lvert E \rvert)$ time.
Moreover, if we precompute and store $\sum_{u \in J_{l}} e^{\theta_{u}}$ for all
metric-balls, then the second term can be computed in
$\mathcal{O}(\lvert V \rvert \log \lvert V \rvert)$ time.
Therefore, the overall cost of evaluating the core score derivatives for all the
vertices is $\mathcal{O}(\lvert V \rvert \log \lvert V \rvert + \lvert E \rvert)$.
Furthermore, we can also approximate the derivative of the objective function
with respect to $\epsilon$:
\begin{align*}
  \textstyle \frac{\partial \Omega}{\partial \epsilon} &= \textstyle \sum_{u<v \in E} -\log K_{uv} + \sum_{u<v} \frac{e^{\theta_{u}+\theta_{v}} \log K_{uv}}{e^{\theta_{u}+\theta_{v}} + K_{uv}^{\epsilon}}  \nonumber \\
& \textstyle \sum_{u<v \in E} -\log K_{uv} + \sum_{l} \sum_{\substack{I < J \in B_{l}\\ p(I) = p(J)}} \sum_{u \in I, v \in J} \frac{e^{\theta_{u} + \theta_{v}} \log K_{uv}}{e^{\theta_{u} + \theta_{v}} + K_{uv}^{\epsilon}}.
\end{align*}
The first term in the last equation can be calculated in
$\mathcal{O}(\lvert E \rvert)$ time, while the second term can be evaluated in
$\mathcal{O}(\lvert V \rvert \log \lvert V \rvert)$ following the same scheme as the objective function,
\begin{align}
\sum_{\substack{u \in I \\ v \in J}} \frac{e^{\theta_{u} + \theta_{v}} \log K_{uv}}{e^{\theta_{u} + \theta_{v}} + K_{uv}^{\epsilon}} \approx \frac{\sum_{u \in I} e^{\theta_{u}} \cdot \sum_{v \in J} e^{\theta_{v}} \cdot \log K_{IJ}}{\frac{1}{\lvert I \rvert}\sum_{u \in I} e^{\theta_{u}} \cdot \frac{1}{\lvert J \rvert}\sum_{v \in J} e^{\theta_{v}} + K_{IJ}^{\epsilon}}.
\label{eq:domega_depsilon_fmm_approx}
\end{align}
Unlike the FMM-style approximation for the objective function
(\cref{eq:omega_fmm_approx}), the denominators in the expression of the
objective function gradient
(\cref{eq:domega_dtheta_fmm,eq:domega_depsilon_fmm_approx}) are not metric
kernels due to the extra terms that involve the vertex core scores.
However, when the given network is sparse and the fitted core scores are small,
the metric kernels $K_{I_{l}J_{l}}^{\epsilon}$ and $K_{IJ}^{\epsilon}$ dominate the
denominators of \cref{eq:domega_dtheta_fmm} and
\cref{eq:domega_depsilon_fmm_approx}, making them good approximations for metric
kernels.


The FMM-style algorithm enables efficient model inference, and the learned core scores
are accurate enough for data mining purposes, which we show in \cref{sec:data}.
Furthermore, we also want to generate random spatial networks, which naively
takes $\mathcal{O}(\lvert V \rvert^{2})$ time. We show how to accelerate this process in the
next section.

\subsection{Efficient Random Network Generation \label{subsec:generation_fmm}}
Given the core scores, we can naively sample a random networks by flipping a biased
coin for each pair of vertices.
However, this process is inefficient when the output network is sparse.
To address this problem, we develop an efficient algorithm to hierarchically
sample the edges. The key idea is to sample edges between pairs of metric-balls
instead of pairs of vertices.

Our algorithm proceeds as follows.
First, we recursively partition the vertices to fit into a metric-tree.
Second, at each level of the tree, for every pair of sibling metric-balls
$I,J$, we compute the expected number of edges $n_{IJ}$ that has one end in each
of $I$ and $J$:
\begin{align}
n_{IJ} \approx \frac{\sum_{u \in I} e^{\theta_{u}} \cdot \sum_{v \in J} e^{\theta_{v}}}{\frac{1}{\lvert I \rvert}\sum_{u \in I} e^{\theta_{u}} \cdot \frac{1}{\lvert J \rvert}\sum_{v \in J} e^{\theta_{v}} + K_{IJ}^{\epsilon}}.
\label{eq:nIJ}
\end{align}
Third, we determine the edges between every pair of sibling metric-balls $I,J$
by sampling $n_{IJ}$ vertices independently from $I$ with probability
$\rho_{u} \propto \ffrac{e^{\theta_{u}}}{\left[\frac{e^{\theta_{u}}}{\lvert J \rvert} \sum_{v \in J} e^{\theta_{v}} + K_{IJ}^{\epsilon}\right]}$,
and $n_{IJ}$ vertices independently from $J$ with probability
$\rho_{v} \propto \ffrac{e^{\theta_{v}}}{\left[\frac{e^{\theta_{v}}}{\lvert I \rvert} \sum_{u \in I} e^{\theta_{u}} + K_{IJ}^{\epsilon}\right]}$.
Finally, we pair samples from $I$ and $J$ sequentially.
The cost for sampling vertices with non-uniform probability in
the third step is $\mathcal{O}(\lvert I \rvert + \lvert J \rvert + n_{IJ} \log n_{IJ})$.
Assuming the generated network is sparse and the number of edges grows linearly
with the number of vertices ($n_{IJ} \propto \lvert I \rvert+\lvert J \rvert$), the overall cost
is $\mathcal{O}\left(\lvert V \rvert (\log \lvert V \rvert)^{2}\right)$.

Creating edges by pairing up independently sampled vertices is called
``ball-dropping'' in Erd\H{o}s-R\'{e}nyi sampling~\cite{Ramani_2017}.
In our model, the number of edges between metric-balls $I$ and $J$ does
not have a closed-formed probability distribution, and our algorithm is only
taking the mean of that distribution by sampling $n_{IJ}$ edges.
Therefore, our fast sampling scheme is only an approximation without any
theoretical guarantees.
However, as we show in the next section, this approximation empirically
preserves network properties.


\section{Methodological experiments}\label{sec:alg-experiments}
In this section, we first numerically validate our approximations and then show
that our model achieves much higher likelihood than competing generative models
for core-periphery structure.
Section 5 then explores data mining tasks aided by our model.
\subsection{Approximation validation}\label{subsec:experiment_fmm}
In the previous section, we made a number of approximations in our methods.
We now show that these approximations indeed have small error.
To optimize the objective function, we use the limited-memory BFGS (L-BFGS)
algorithm from the \texttt{Optim.jl} package~\cite{Mogensen_2018}.
We also use a diagonal preconditioner in case the approximated Hessian used
within L-BFGS becomes ill-conditioned.
We construct the metric-tree with the \texttt{NearestNeighbors.jl} package.
Finally, we set the tolerance parameters to $\delta_{1}=2.0$ and $\delta_{2}=0.2$.%
\footnote{Our software is available at \url{https://github.com/000Justin000/spatial_core_periphery}.}

\begin{figure}[tb]
  \centering
  \phantomsubfigure{fig:celegans_accuracy_A}
  \phantomsubfigure{fig:celegans_accuracy_B}  
\includegraphics[width=0.98\linewidth]{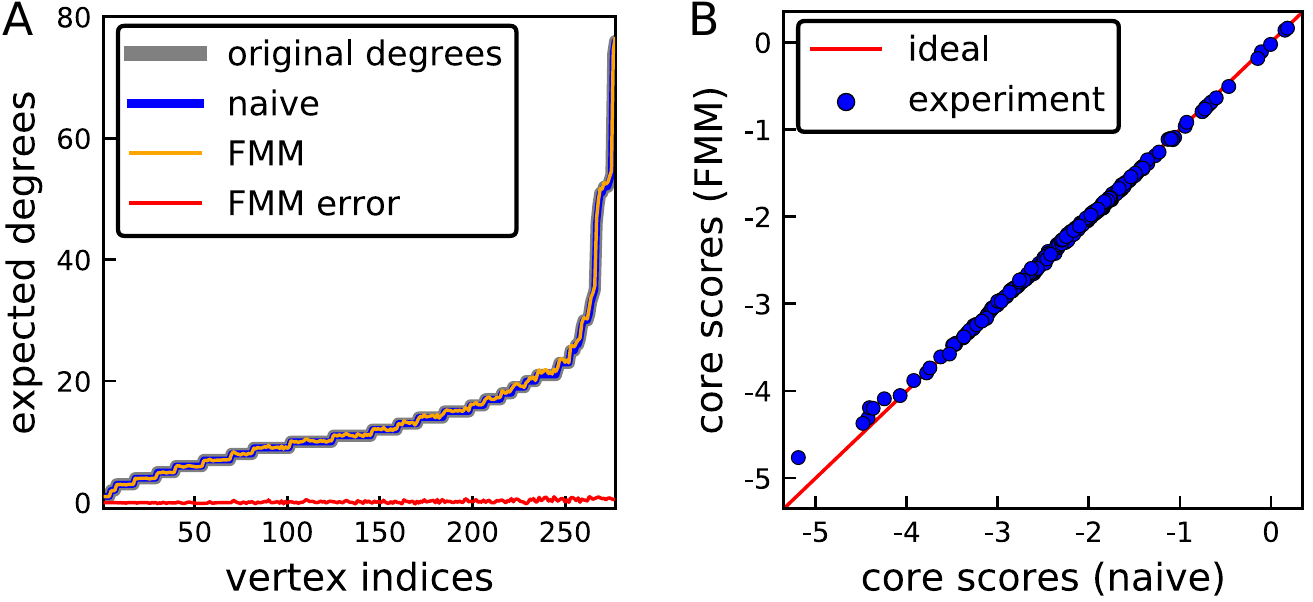}
\caption{%
(A)
Agreement between the naive (blue) and fast (orange) algorithms
on the expected vertex degrees. The error of the FMM-style approximation is in red.
We also include the vertex degrees in the original network (gray) as numerical
validation for \cref{thm:degree}.
(B)
Correlation between the learned core scores from the naive algorithm and the FMM
algorithm. The Pearson correlation coefficient is 0.999.
}
\label{fig:celegans_accuracy}
\end{figure}
\xhdr{Accuracy in evaluating the objective function and derivative}
To test our fast algorithm, we first
learn parameters for the \textit{C. elegans} network from \cref{fig:com_cp}
using the naive algorithm and the Euclidean distance kernel.
Next, we evaluate $\Omega$ and the gradient using the same parameters, but this
time using the fast algorithm to sum over the pairwise interactions between
vertices.
The relative difference in $\Omega$ between the naive and the fast algorithms
was less than 1\%.
To evalualuate agreement in the derivative, we compare expected degrees
obtained by the fast algorithm and the naive algorithm and find
them to be nearly identical (\cref{fig:celegans_accuracy_A}).
As a sanity check, the expected vertex degrees computed by both algorithms are
very close to the vertex degrees in the original network, which is a necessary
condition for $\Omega$ to be at a local maximum (\cref{thm:degree}).

To understand how the quality of the approximation depends on the input accuracy parameters, 
we repeat the above experiments using different combinations of $\delta_{1}$ and $\delta_{2}$.
We characterize the approximation error with the root mean square error in expected
degrees between the fast algorithm and the naive algorithm (\cref{fig:celegans_gen_analysisA})
and report dependence of the running time on the accuracy parameters (\cref{fig:celegans_gen_analysisB}).
Both accuracy and running time have a stronger dependency on $\delta_{2}$ as compared to $\delta_{1}$.
This reason is that when our algorithm determines the radius of the parent
metric-ball, it assumes the worst-case condition and chooses a large value.
Therefore in most cases, $\delta_{1}=1.0$ could already guarantee that all the vertices between two metric-balls
are well-separated.

Next, we use our fast algorithm to learn the model parameters.
The learned vertex core scores with the naive and the fast algorithms had Pearson
correlation of $0.999$ and empirically look similar
(\cref{fig:celegans_accuracy_B}).
Furthermore, the learned $\epsilon$ by the naive algorithm and the FMM algorithm
were $0.499$ and $0.506$, respectively.

\begin{figure}[tb]
\phantomsubfigure{fig:celegans_gen_analysisA}
\phantomsubfigure{fig:celegans_gen_analysisB}
\centering
\includegraphics[width=0.98\linewidth]{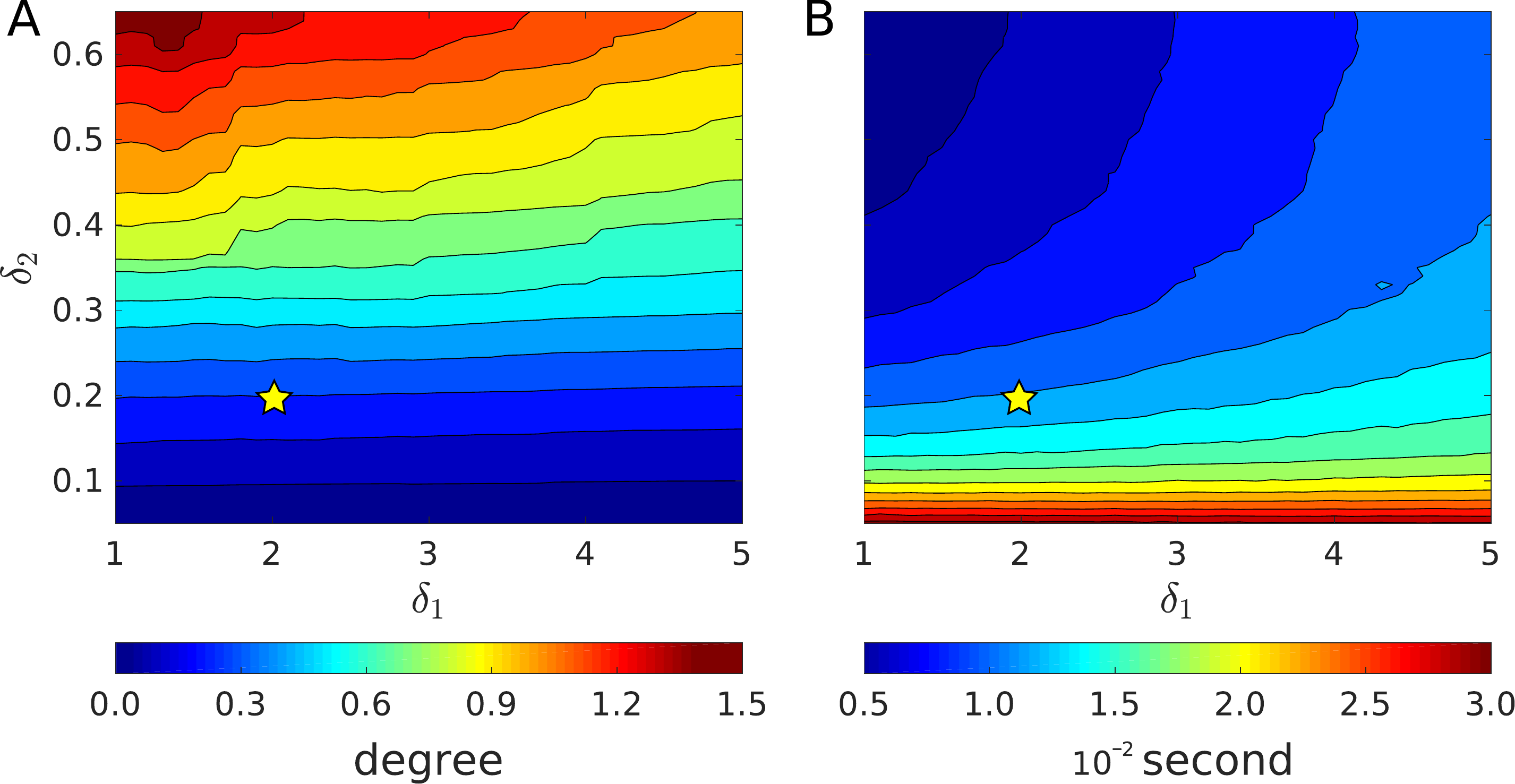}
\caption{
Performance dependence on accuracy parameters on the {\em C. elegans} network.
The yellow stars represent the location of default parameters.
(A)
Root mean square error in expected degrees.
(B)
Running time per gradient evaluation.
}
\label{fig:celegans_tradeoff}
\end{figure}

\begin{figure}[tb]
\phantomsubfigure{fig:celegans_gen_analysisA}
\phantomsubfigure{fig:celegans_gen_analysisB}
\centering
\includegraphics[width=0.95\linewidth]{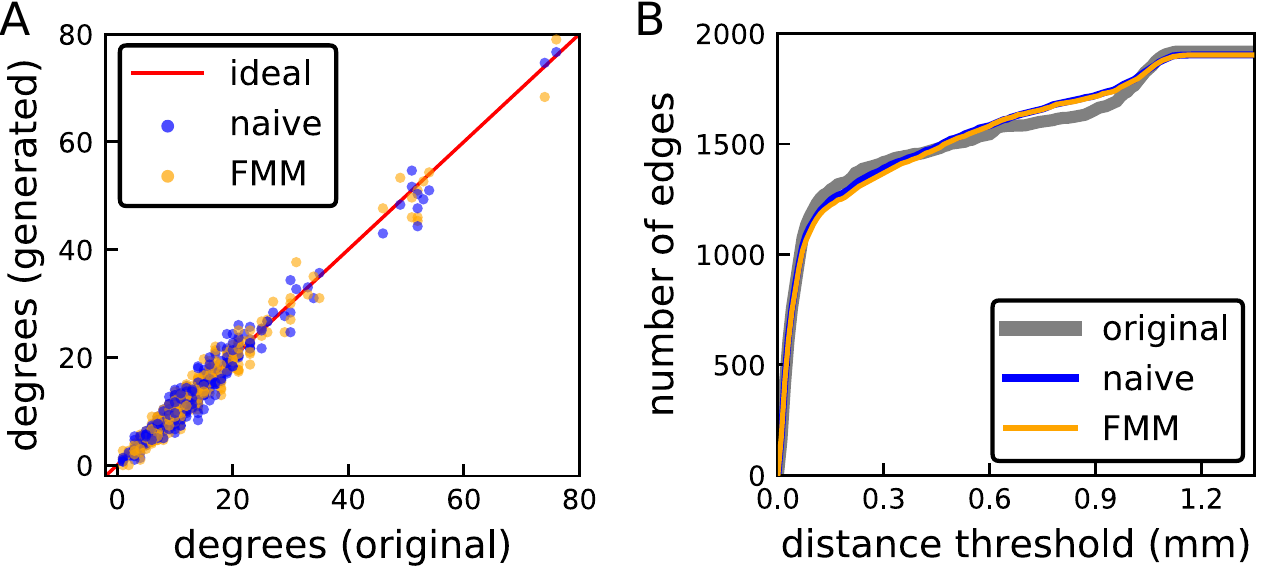}
\caption{
(A) Correlation of vertex degrees in \emph{C. elegans} and the random networks
generated by the naive and fast algorithm; the Pearson correlation coefficients
are 0.983 and 0.981, respectively.
(B) Number of edges whose kernel distance is below a distance threshold.
The agreement numerically validates our fast sampling algorithm
as well as our arguments in \cref{sec:algo} on preserving distances.
}
\label{fig:celegans_gen_analysis}
\end{figure}
\xhdr{Quality of generated random networks}
Using the model parameters learned by the naive algorithm, we now confirm that random networks
generated by the naive sampling algorithm and the fast sampling algorithm are
similar.
We first consider the degree distribution.
For both the naive algorithm and the fast algorithms, we compare the mean
degrees over three sampled networks to the degree in the original network
(\cref{fig:celegans_gen_analysisA}).
Indeed, the degrees in the random networks generated by both
algorithms are highly correlated with that in the original network.

Next, we consider the edge length distribution given by the kernel function.
We argued that the expected log-GMEL in the generated random networks equals
that in the original network at any local maximum of the likelihood function.
Indeed, the GMEL in \textit{C. elegans} is $0.079$mm, while the
GMEL in the random networks generated by the naive algorithm and the fast
algorithm (averaged over three samples) are $0.078$mm and $0.079$mm.
For a more detailed analysis, we picked equispaced distance
thresholds from $0.0$mm to $1.4$mm and counted the number of edges in the
network whose length is smaller than each threshold.
The counts of the original network, the mean of three naive random samples,
and the mean of three samples with our fast algorithm are nearly identical
(\cref{fig:celegans_gen_analysisB}).

\begin{figure}[tb]
\centering
\includegraphics[width=1.00\linewidth]{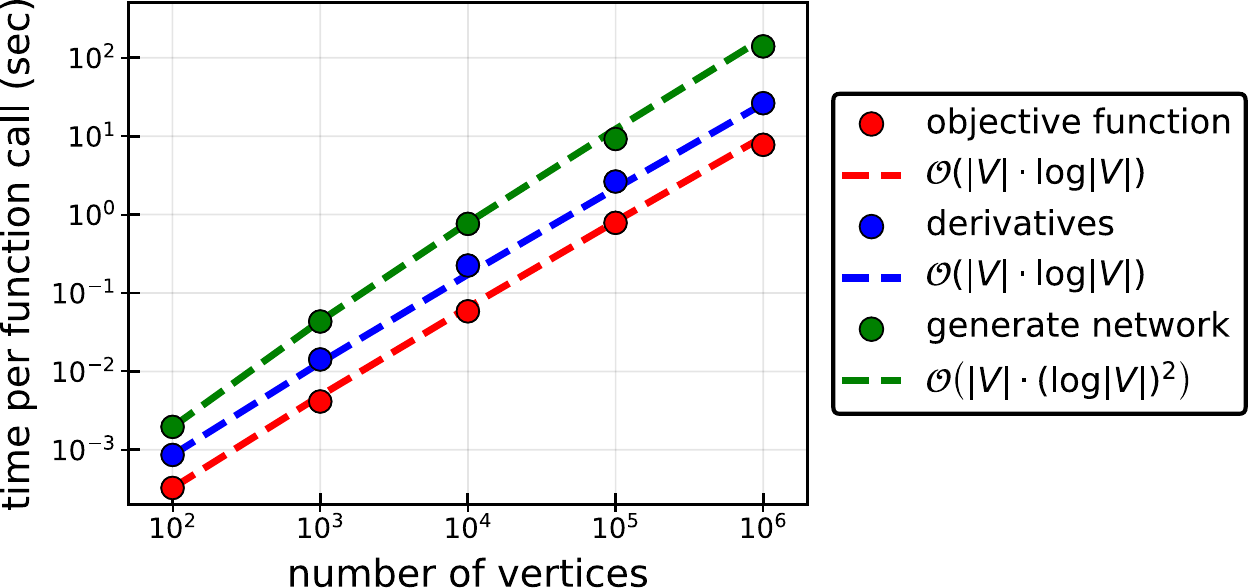}
\caption{
Scalability of our FMM-style algorithms using synthetic networks,
where the average degree in each network is 10.
Observed timings are scattered with circles, and ideal efficiencies are
plotted in dashed lines.
}
\label{fig:timings}
\end{figure}

\xhdr{Scalability of the FMM-style algorithm}
Finally, we validate the computational complexity of our proposed methods.
We run our algorithms on synthetic 2-block core-periphery
networks where vertex coordinates are randomly generated in the
2-dimensional unit square.
We choose 5\% of the vertices in the networks as the core, and their core score
(denoted as $\theta_{c}$) is set to be $1.0$ unit larger than the core score of
the periphery vertices (denoted as $\theta_{p}$).
The core scores $\theta_{c}, \theta_{p}$ in this family of networks are chosen
so that the number of edges grows linearly with the number of
vertices (in the five networks, $\theta_{c} =$ $-1.25$, $-2.77$, $-4.13$, $-5.43$, $-6.69$).
%
\cref{fig:timings} shows that the empirical running time follows the theoretical analysis.

\subsection{Likelihood comparison}\label{subsec:results_likelihood}
We now demonstrate the quality of our model by comparing its optimized
log-likelihood with the only other two generative models for
core-periphery structure. The first is a 2-block stochastic block model with a
belief propagation learning algorithm that explicitly incorporates
core-periphery structure (henceforth, SBM-CP)~\cite{Zhang_2015}.
Since the belief propagation algorithm is sensitive to the initial conditions,
we perform $5$ independent runs of the algorithm from different random initial
conditions and record the best result.
The second model defines the edge probability as a logistic function
of the ``centrality rank'' of the incident vertices (henceforth, logistic-CP)~\cite{Tudisco_2018}.
For this model, we tune the hyperparameters $s,t$ in the logistic function with grid search
and record the best result.
%

%
\begin{table}[tb]
\setlength{\tabcolsep}{2pt}
\centering
\caption{Number of nodes ($n$), number of edges ($m$), and optimized
  log-likelihood of our model, SBM-CP and logistic-CP. In all datasets, our
  model has a larger likelihood. (Results for logistic-CP are only available
  for connected graphs.)}
\label{tab:log_likelihood}
\begin{tabular}{ l l l c c c}
\toprule
& & & \multicolumn{3}{c}{Log likelihood} \\
\cmidrule(r){4-6} 
Dataset                 & $n$       & $m$       & our model         & SBM-CP            & logistic-CP  \\
\midrule
{\small {\em C. elegans}}       & 279       & 1.9K     & -$6.3 \cdot 10^{3}$ & $-7.1 \cdot 10^{3}$ & $-7.0 \cdot 10^{3}$ \\
{\small London Under.}     & 315       & 370       & $-6.0 \cdot 10^{2}$ & $-2.2 \cdot 10^{3}$ & $-2.1 \cdot 10^{3}$ \\
{\small Pv\_M\_I\_U\_N\_42d\_1} & 2.4k     & 2.6K     & $-6.4 \cdot 10^{3}$ & $-2.1 \cdot 10^{4}$ & ---                  \\
{\small OpenFlights}            & 7.2K     & 18.6K    & $-4.7 \cdot 10^{4}$ & $-1.1 \cdot 10^{5}$ & ---                  \\
{\small Brightkite}             & 50.7K    & 194K   & $-1.3 \cdot 10^{6}$ & $-1.9 \cdot 10^{6}$ & ---                  \\
{\small LiveJournal}            & 1.16M & 7.19M & $-7.5 \cdot 10^{7}$ & $-8.9 \cdot 10^{7}$ & ---                  \\
\bottomrule
\end{tabular}
\end{table}

We learn the parameters of our model, SBM-CP and logistic-CP on six datasets (basic summary statistics
are listed in \cref{tab:log_likelihood}):
\newline
(i) The neural network of the nematode worm \emph{C. elegans} from \cref{fig:com_cp}.
\newline
(ii) The network of the London underground transportation system, where
vertices are tube stations and edges connect stations.%
\footnote{The network was derived based on a similar one of Rombach et
  al.~\cite{Rombach-2014-CP}.  Vertex coordinates were collected from
  Wikipedia.}
\newline
(iii) The fungal network Pv\_M\_I\_U\_N\_42d\_1 constructed from
a laboratory experiment, where each hypha (tubular cell) is a
vertex, and each cord transporting nutrients between two hyphae is an
edge~\cite{Lee_2017}. The fungus grows on a 2-dimensional plane.
\newline
(iv)
The airline transportation network from OpenFlights.%
\footnote{Data collected from \url{https://openflights.org/data.html\#route}.}
Vertices are airports with latitude and longitude spatial locations,
and edges correspond to available direct flights.
\newline
(v) The Brightkite social network~\cite{cho2011friendship}.
Vertices are users and spatial locations come from
the user's most recent check-in.
\newline
(vi) The LiveJournal social network~\cite{Liben-Nowell_2005}. Vertices are
bloggers and the spatial locations are given by the city listed in the user's
profile. Edges are friendships listed in a user's profile.

A small amount of noise was added to locations in the Brightkite and LiveJournal
datasets to avoid having vertices in the same location.
For the OpenFlights, Brightkite, and LiveJournal datasets, spatial positions are
latitude and longitude, so we use the great circle distance
kernel. For \emph{C. elegans} and the fungal network, we use Euclidean
distance. In the London Underground dataset, tube stations are almost always
connected to their closest spatial neighbors due to construction cost of
railways. Therefore, we choose a symmetric version of rank
distance~\cite{Liben-Nowell_2005} as the kernel.
\cref{tab:log_likelihood} lists the optimized log-likelihoods of our model, SBM-CP and logistic-CP;
our model always has substantially higher likelihood.



\section{Data mining experiments}\label{sec:data}
Now that we have addressed the accuracy, scalability, and efficacy of our
algorithms and model, we use the output of our model to study real-world spatial
networks.
We find that the learned core scores from our model are good predictors
for traffic in an airport network and for the classification of fungal networks.

\subsection{Case study I: Predicting enplanement}\label{sec:enplanement}
\begin{figure}[tb]
\centering
\includegraphics[width=0.86\linewidth]{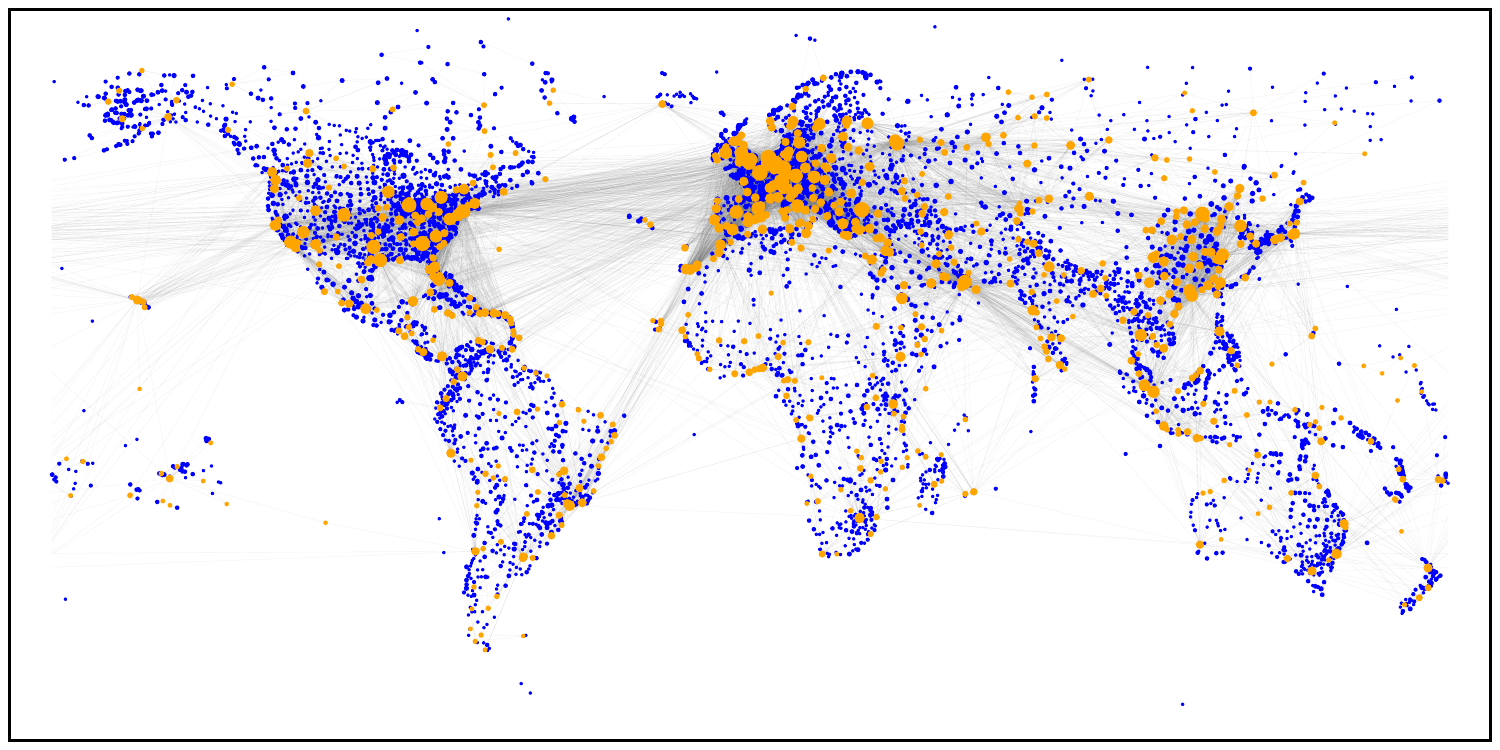}
\caption{
Vertex core scores in the airline network.
The top 10\% of the vertices with highest core scores are colored in orange,
while the rest are colored in blue.
The radius of a vertex is proportional to the square root of its degree.
}
\label{fig:openflight}
\end{figure}

\begin{figure}[tb]
\centering
\includegraphics[width=1.0\linewidth]{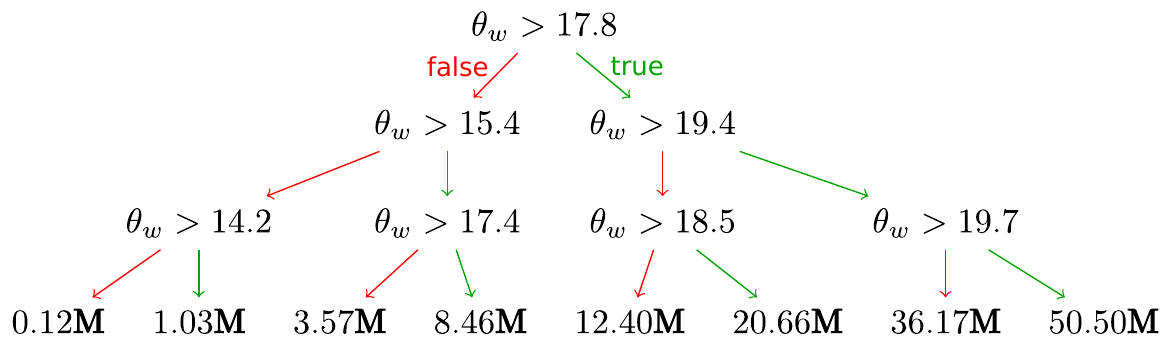}
\caption{%
Top three levels of the decision tree for predicting airport enplanement using
vertex core scores.
The leaf nodes in the decision tree is the average enplanement (in millions)
among airports with core scores in the given range.
}
\label{fig:decision_tree}
\end{figure}

We first analyze the OpenFlights network introduced in
\cref{subsec:results_likelihood}. Recall that this network has vertices
representing airports and edges representing direct flights between airports.
Given the network and vertex coordinates, our goal is to predict the total number
of passengers boarding (enplanement) at each airport in the year 2017 using
vertex core scores.
To this end, we first compute the vertex core scores using the FMM-style
algorithm with the great-circle distance kernel (\cref{fig:openflight}) and
then we use a decision tree to correlate the vertex core scores with
enplanement (\cref{fig:decision_tree})
 
We obtained the enplanement metadata for $447$ airports in the United States.%
\footnote{Data collected from \url{https://www.faa.gov/airports/planning_capacity}.}
We first split the airports randomly into a training set of $357$ airports ($80\%$)
and test set of $90$ airports ($20\%$); after, we build a decision tree using the
training set and test its prediction accuracy on the testing set.
\Cref{fig:decision_tree} shows an instance of the decision tree we build using the
training set.

We average the prediction accuracy of decision trees over $10$ random split of
the training and testing set, and the mean coefficient of determination ($R^{2}$)
between ground truth and predicted enplanement is $0.846$.
Analogous experiments are repeated using vertex degrees, betweenness centrality
(BC)~\cite{Freeman_1977}, closeness centrality (CC)~\cite{Bavelas_1950},
eigenvector centrality (EC)~\cite{Newman_2003} or PageRank
(PR)~\cite{Pageetal_1998} instead of the core score as the independent variable
(\cref{tab:openflight_accuracy}).
The vertex core scores outperforms other centrality measures in characterizing
airport enplanement.
This high accuracy indicates that our model effectively utilizes spatial information.

\begin{table}[tb]
\setlength{\tabcolsep}{5pt} 
\centering
\caption{The $R^{2}$ values between ground truth and predicted enplanement with different features.
  Our learned core scores have the largest value.}
\label{tab:openflight_accuracy}
\begin{tabular}{r @{\qquad} c c c c c c}
\toprule
        &   degree &       BC &       CC &      EC &       PR & core score \\
\midrule
$R^{2}$ &  $0.762$ &  $0.293$ &  $0.663$ & $0.542$ &  $0.637$ &    $\textbf{0.846}$ \\
\bottomrule
\end{tabular}
\end{table}

\subsection{Case study II: Classifying fungal networks}
In this case study, we use core scores to predict types of fungal networks
(the network examined in \cref{subsec:results_likelihood} is
one such network).
In these networks, vertices are hyphae in 2-dimensional Euclidean space
and edges are cords transporting nutrients between two hyphae.
Based on the species and growing conditions, the fungal networks
were categorized into $15$ different classes~\cite{Lee_2017}.
Here, we use core scores as features to predict the class label of each network.

\begin{figure}[tb]
\centering
\includegraphics[width=1.0\linewidth]{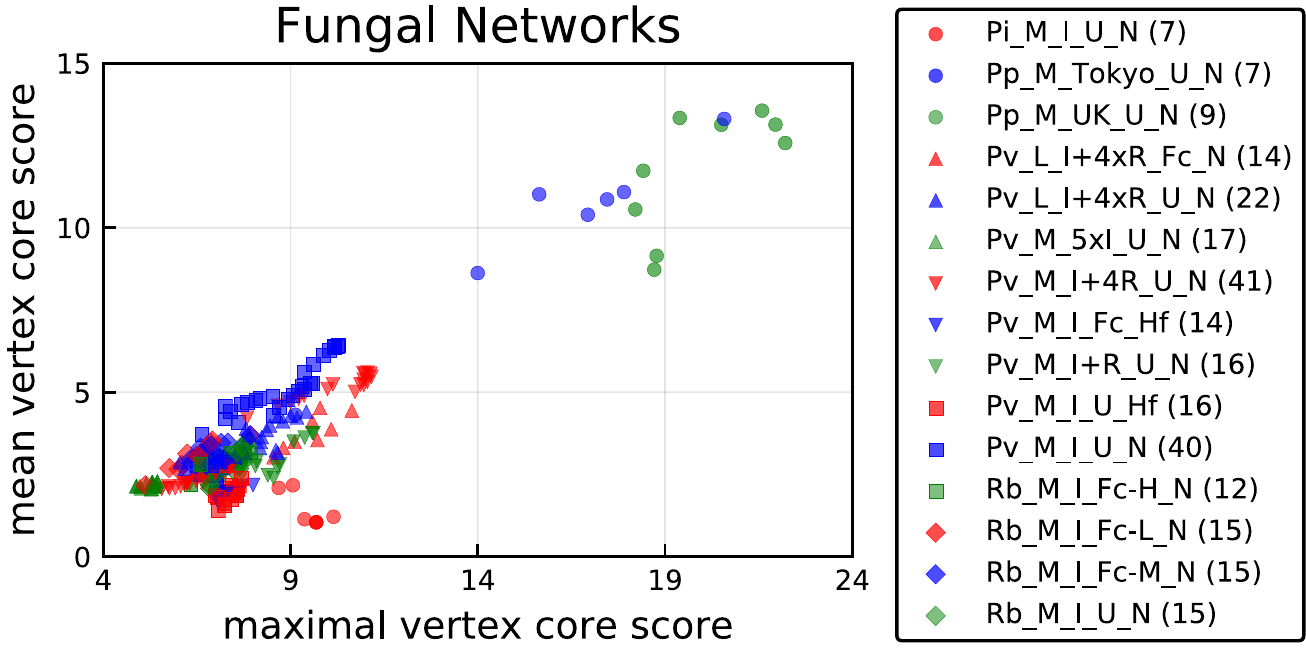}
\caption{%
Maximal and mean vertex core scores for 260 fungal networks.
The integer in the parentheses after each class label is the number of networks
in that class.
The plot shows some clustering of class type based on these features.
We use these features along with a couple others derived from the learned core
scores to train a logistic regression classifier with good performance
(\cref{tab:fungal_accuracy}).
}
\label{fig:fungal_networks}
\end{figure}

First, we learned the vertex core scores for every fungal network using our
model with the Euclidean distance kernel and the FMM-style algorithm.
Then, for each network, we created a 4-element feature vector using the maximum, mean, and
standard deviation of the core scores, along with the number of vertices.
We trained a logistic regression model with these features (and an intercept term)
and evaluated prediction accuracy with 5-fold cross validation.
We repeated the experiment using the centrality measures from \cref{sec:enplanement}.
\Cref{tab:fungal_accuracy} shows that core scores from our model are the best predictors.
As evidence for why this might be true, \cref{fig:fungal_networks} plots
the mean and maximal core score for each network, along with the class labels.
We see class separation with these two features.

Furthermore, we use a logistic regression model by concatenating
information from vertex degrees and core scores as features.
In other words, the predictors are the maximum, mean and standard deviation of
vertex degrees; the maximum, mean, and standard deviation of vertex core scores;
and the number of vertices.
We observe a large increase in prediction accuracy from $43.5\%$ to $65.6\%$.
This result shows that vertex core scores captures different information than
degrees in the fungal networks.

\begin{table}[tb]
\setlength{\tabcolsep}{4pt} 
\centering
\caption{Cross validation accuracy of fungal networks classification using different features.
Our learned core scores give the best prediction. ``Random'' is random guessing.}
\label{tab:fungal_accuracy}
\begin{tabular}{c c c c  c c c}
\toprule
   random  &    degree &        BC &        CC &       EC &       PR & core score \\
\midrule
6.7\%  &  38.0\% & 23.7\% & 19.0\% & 20.2\% & 18.6\% & \textbf{43.5\%} \\
\bottomrule
\end{tabular}
\end{table}

\section{Additional related work \label{sec:review}}
The idea of core-periphery structure has a long history in the theory of social
networks~\cite{Burt-1976-Positions,Breiger-1981-structures,Borgatti_2000},
where the core arises due to differential status.
Borgatti and Everett developed the first significant computational approach to identify
core-periphery structure~\cite{Borgatti_2000}.
Since then, several methods have been deisgned to find core-periphery structure,
based on paths~\cite{Lee_2014,Gamble_2016},
vertex covers~\cite{benson2018found},
spectral information~\cite{Cucuringu_2016,Ma_2018},
$k$-cores and generalizations~\cite{dumba2018uncovering,sariyuce2015finding},
hand-crafted ``core quality'' objective functions~\cite{Rombach-2014-CP,Rombach_2017,Kojaku_2018},
and user studies~\cite{joblin2017classifying}.
Such methods are typically analyzed via synthetic benchmarks, density measures,
or network visualizations.
The key difference with our work is that we propose a \emph{generative model},
whereas prior work performs \emph{post hoc identification} of core-periphery
structure using the network topology.
The only other generative models are due to Zhang et al.~\cite{Zhang_2015} and
Tudisco and Higham~\cite{Tudisco_2018} against which we compared in
\cref{subsec:results_likelihood} (unlike our approach, these methods do not use spatial information).

Network centrality captures similar properties to core-periphery structure.
Indeed, closeness centrality is used to detect core vertices in
networks~\cite{Holme_2005,Silva_2008}, and centrality measures are
baselines for core-periphery identification
measures~\cite{Rombach_2017} (see also our experiments in \cref{sec:data}).
A subtlety is that network centrality can be a consequence of core-periphery
dynamics, but centrality in and of itself does not lead to a \emph{reason} for
core-periphery structure.
This is one reason why we developed a generative model in this paper.

\section{Discussion}
We have developed a random network model for core-periphery structure, where
each vertex has a real-valued core score.
We focused our analysis on spatial data, which connects to the small-world model
where edge probabilities are inversely correlated with distance.
The spatial structure enabled us to develop fast algorithms for both
gradient-based inference and network sampling.
We showed both theoretically and numerically that our model preserves the
expected degree of each vertex as well as the aggregated log-distance.
Our model can be an alternative to the Chung-Lu model, even if there is no
spatial metadata.
Furthermore, our model does not need to learn from a network---given a
prescribed set of core scores and a kernel function, we can generate networks.

In terms of likelihood, our model out-performs the only other generative models
for core-periphery structure (SBM-CP and logistic-CP).
The basic version of our model can be thought of as a continuous relaxation of
SBM-CP, since the core scores permit a continuum of edge probabilities.
Finally, we also demonstrated that the learned core scores are useful vertex
features for downstream network analysis and machine learning tasks in two very
different complex systems (airport traffic and fungal growth), which provides
evidence that our model could be incorporated into a wide area of application
domains.

\begin{acks}
We thank David Liben-Nowell for providing the LiveJournal data. 
This research was supported in part by NSF Award DMS-1830274.
\end{acks}

\bibliography{main}
\bibliographystyle{ieeetr}

\end{document}